# Earthquake Networks, Complex

SUMIYOSHI ABE[a,b] and NORIKAZU SUZUKI[c]

[a]Department of Physical Engineering, Mie University, Tsu, Mie 514-8507, Japan

[b]Institut Supérieur des Matériaux et Mécaniques Avancés, 44 F. A. Bartholdi, 72000 Le Mans, France

[c]College of Science and Technology, Nihon University, Chiba 274-8501, Japan

**Article Outline**

**Glossary**

**Network or Graph**

A network (or a graph) [1] consists of vertices (or nodes) and edges (or links) connecting them. In general, a network contains loops (i.e., edges with both ends attached to the same vertices) and multiple edges (i.e., edges more than one that connect two different vertices). If edges have their directions, such a network is called directed.

A simple graph is a network, in which loops are removed and each multiple edge is replaced by a single edge. In a stochastic network, each connection is inherently probabilistic. A classical random graph is a simple example, in which each two vertices are connected by an edge with probability $p$ and unconnected with probability $1-p$ $(0 < p < 1)$.

**Connectivity Distribution or Degree Distribution**

The connectivity distribution (or the degree distribution), $P(k)$, is the probability of finding vertices with $k$ edges in a stochastic network. In a directed network, the number of incoming/outgoing edges is called the in-degree/out-degree. Connectivity of a classical random graph obeys the Poissonian distribution in the limit of the large number of vertices [2-4], $P(k) = e^{-\lambda} \lambda^k / k!$ ($\lambda$: a positive parameter, $k = 0, 1, 2, ...$), whereas a scale-free network [2-4,7] has a power-law shape, $P(k) \sim k^{-\gamma}$ ($\gamma$: a positive exponent), for large $k$.

**Preferential Attachment Rule**

This is a concept relevant to a growing network, in which the number of vertices increases. Preferential attachment [2-4,7] implies that a newly created vertex tends to link to pre-existing vertices with the probability $\Pi(k_i) = k_i / \sum_j k_j$, where $k_i$ stands for the connectivity of the $i$th vertex. That is, larger the connectivity of a vertex is, higher the probability of getting linked to a new vertex is.

**Clustering Coefficient**

The clustering coefficient [8] is a quantity characterizing an undirected simple graph. It quantifies the adjacency of two neighboring vertices of a given vertex, i.e., the tendency of two neighboring vertices of a given vertex to be connected to each other. Mathematically, it is defined as follows. Assume the $i$th vertex to have $k_i$ neighboring vertices. There can exist at most $k_i(k_i - 1)/2$ edges between the neighbors. Define $c_i$ as the ratio

$$c_i = \frac{\text{actual number of edges between the neighbors of the } i\text{th vertex}}{k_i(k_i - 1)/2}. \quad (1)$$

Then, the clustering coefficient is given by the average of this quantity over the network:

$$C = \frac{1}{N} \sum_{i=1}^{N} c_i, \quad (2)$$

where $N$ is the total number of vertices contained in the network. The value of the clustering coefficient of a random graph, $C_{\text{random}}$, is much smaller than unity, whereas a small-world network has a large value of $C$ which is much larger than $C_{\text{random}}$.

**Hierarchical Organization**

Many complex networks are structurally modular, that is, they are composed of groups of vertices that are highly interconnected to each other but weakly connected to outside groups. This hierarchical structure [9] can conveniently be characterized by the clustering coefficient at each value of connectivity, $c(k)$, which is defined by

$$c(k) = \frac{1}{N P_{\text{SG}}(k)} \sum_{i=1}^{N} c_i \, \delta_{k_i, k}, \quad (3)$$

where $c_i$ is given by Eq. (1), $N$ the total number of vertices, and $P_{\text{SG}}(k)$ the connectivity distribution of an undirected simple graph. Its average is the clustering coefficient in Eq. (2): $C = \sum_k c(k) P_{\text{SG}}(k)$. A network is said to be hierarchically organized if $c(k)$ varies with respect to $k$, typically due to a power law, $c(k) \sim k^{-\beta}$, with a positive exponent $\beta$.

**Assortative Mixing and Disassortative Mixing**

Consider the conditional probability, $P(k'|k)$, of finding a vertex with connectivity $k'$ linked to a given vertex with connectivity $k$. Then, the nearest-neighbor average connectivity of vertices with connectivity $k$ is defined by [3,10,11]

$$\bar{k}_{nn}(k) = \sum_{k'} k' \, P(k'|k). \quad (4)$$

If $\bar{k}_{nn}(k)$ increases/decreases with respect to $k$, mixing is termed assortative/disassortative. A simple model of growth with preferential attachment is known to possess no mixing. That is, $\bar{k}_{nn}(k)$ does not depend on $k$.

The above-mentioned linking tendency can be quantified by the correlation coefficient [12] defined as follows. Let $e_{kl}\,(=e_{lk})$ be the joint probability distribution for an edge to link with a vertex with connectivity $k$ at one end and a vertex with connectivity $l$ at the other. Calculate its marginal, $q_k = \sum_l e_{kl}$. Then, the correlation coefficient is given by

$$r = \frac{1}{\sigma_q^2} \sum_{k,l} kl\,(e_{kl} - q_k q_l), \tag{5}$$

where $\sigma_q^2 = \sum_k k^2 q_k - (\sum_k k q_k)^2$ stands for the variance of $q_k$. $r \in [-1, 1]$, and if $r$ is positive/negative, mixing is assortative/disassortative [3,12].

I. **Definition of the Subject and Its Importance**

Complexity is an emergent collective property, which is hardly understood by the traditional approach in natural science based on reductionism. Correlation between elements in a complex system is strong, no matter how largely they are separated both spatially and temporally, therefore it is essential to treat such a system in a holistic manner, in general.

Although it is generally assumed that seismicity is an example of complex phenomena, it is actually nontrivial to see how and in what sense it is complex. This point may also be related to the question of primary importance why it is so difficult to predict earthquakes.

Development of the theory of complex networks turns out to offer a peculiar perspective on this point. Construction of a complex earthquake network proposed here consists of mapping seismic data to a growing stochastic graph. This graph, or network, turns out to exhibit a number of remarkable behaviors both physically and mathematically, which are in common with many other complex systems. The scale-free and small-world natures are typical examples. In this way, one will be able to obtain a novel viewpoint of seismicity.

## II.  Introduction

Seismicity is a field-theoretical phenomenon. Released energy of each earthquake may be regarded as a field amplitude defined at a discrete spacetime point. However, in contrast to a familiar field theory such as the electromagnetic theory, both amplitudes and locations are intrinsically probabilistic. The fault distribution may geometrically be fractal [13], and the stress distribution superposed upon it often has a complex landscape. Accordingly, seismicity is characterized by extremely rich phenomenology, which attracts physicists' attention from the viewpoint of science of complex systems.

There are at least two celebrated empirical laws known in seismology. One is the Gutenberg-Richter law [14], which states that the frequency of earthquakes obeys a power law with respect to released energy. This power-law nature makes it difficult or even meaningless to statistically distinguish earthquakes by their values of magnitude because of the absence of typical energy scales. The other is the Omori law [15], which states that the rate of the frequency of aftershocks following a main shock algebraically decays with respect to time elapsed from the main shock. This slow relaxation reminds one of complex glassy dynamics [16]. Such a viewpoint is supported by the discovery of the aging phenomenon and the scaling law for aftershocks [17].

Another point, which seems less noticed, is that correlation of two successive events is strong, no matter how large their spatial separation is. There is, in fact, an observation [18] that an earthquake can be triggered by a foregoing one more than 1000 km away. The reason why two successive events are indivisibly related can also be found in another observation [19,20] that both spatial distance and time interval between two successive events obey the $q$-exponential distributions in nonextensive statistics [5,6,21], which offers a statistical-mechanical framework for describing complex systems. Thus, the correlation length can be enormously large and long-wave-length modes of seismic waves play an important role. This has a strong similarity to phase transitions and critical phenomena. Accordingly, it may not be appropriate to use spatial windows in analysis of seismicity. Furthermore, all of the data in a relevant area (ideally the whole globe, though still not satisfactorily available) should be treated based on the nonreductionistic standpoint.

The network approach is a powerful tool for analyzing kinematical and dynamical structures of complex systems in a holistic manner. Such a concept was introduced to seismology by the present authors in 2004 [22] in order to represent complexity of seismicity. The procedure described in Section III allows one to map a seismic time series to a growing stochastic network in an unambiguous way. Vertices and edges of such a network correspond to coarse-grained events and event-event correlations, respectively. Yet unknown microscopic dynamics governing event-event correlations and fault-fault interactions are replaced by these edges. Global physical properties of seismicity can then be explored by examining its geometric (e.g., topological etc.) and dynamical features. It turns out that earthquake networks have a number of intriguing properties, some of which are shared by many other natural as well as artificial systems including metabolic networks, food webs, the Internet, the world-wide web, and so on [2-4]. This, in turn, enables seismologists to study seismicity in analogy with such relatively better understood complex systems. Thus, the network approach offers a novel way of analyzing seismic time series and casts fresh light on the physics of earthquakes.

In this article, only the data taken from California is utilized. However, it has been ascertained that the laws and trends discussed here are universal and hold also in other geographical regions including Japan.

**III. Construction of Earthquake Network**

An earthquake network is constructed as follows [22]. A geographical region under consideration is divided into small cubic cells. A cell is regarded as a vertex if earthquakes with any values of magnitude above a certain detection threshold occurred therein. Two successive events define an edge between two vertices. If they occur in the same cell, a loop is attached to that vertex. This procedure enables one to map a given interval of the seismic data to a growing probabilistic graph, which is referred to as an earthquake network (see Fig. 1a).

Several comments are in order. Firstly, this construction contains a single parameter: cell size, which is a scale of coarse graining. Once cell size is fixed, an earthquake network is unambiguously defined. However, since there exist no *a priori* operational rule to determine cell size, it is important to notice how the properties of an earthquake network depend on this parameter. Secondly, as mentioned in Section II, edges and loops efficiently represent event-event correlation. Thirdly, an earthquake network is a directed graph in its nature. Directedness does not bring any difficulties to statistical analysis of connectivity (degree, i.e., the number of edges attached to the vertex under consideration) since, by construction, the in-degree and out-degree are identical for each vertex except the initial and final vertices in analysis. Therefore, the in-degree and out-degree are not distinguished from each other in the analysis of the connectivity distribution (see Sections IV and VII). However, directedness becomes essential when the path length (i.e., the number of edges) between a pair of connected vertices, i.e., the degree of separation between the pair, is considered. This point is explicitly discussed in the analysis of the period distribution in Section VIII. Finally, a full directed earthquake network has to be reduced to a simple undirected graph, when its small-worldness and hierarchical structure are examined (see Sections V and VI). There, loops are removed and each multiple edge is replaced by a single edge (see Fig. 1b). The path length in this case is the smallest value among the possible numbers of edges connecting a pair of vertices.

**IV. Scale-Free Nature of Earthquake Network**

An earthquake network contains some special vertices which have large values of connectivity. Such "hubs" turn out to correspond to cells with main shocks. This is due to a striking fact discovered from real data analysis that aftershocks associated with a main shock tend to return to the locus of the main shock, geographically. This is the primary reason why a vertex containing a main shock becomes a hub. The situation is analogous to the preferential attachment rule for a growing network [2-4]. According to this rule, a newly created vertex tends to be connected to the (already existing) *i*th vertex with connectivity $k_i$ with probability, $\Pi(k_i) = k_i / \sum_j k_j$. It can generate a scale-free network characterized by the power-law connectivity distribution [4,7]:

$$P(k) \sim k^{-\gamma}, \qquad (6)$$

where $\gamma$ is a positive exponent.

In Fig. 2, the connectivity distribution of the full earthquake network with loops and multiple edges is presented [22]. From it, one appreciates that the earthquake network in fact possesses connectivity of the form in Eq. (6) and is therefore scale-free. The smaller the cell size is, the larger the exponent, $\gamma$, is, since the number of vertices with large values of connectivity decreases as cell size becomes smaller. The scale-free nature may be interpreted as follows. As mentioned above, aftershocks associated with a main shock tend to be connected to the vertex of the main shock, satisfying the preferential attachment rule. On the other hand, the Gutenberg-Richter law states that frequency of earthquakes decays slowly as a power law with respect to released energy. This implies that there appear quite a few giant components, and accordingly the network becomes highly inhomogeneous.

## V. Small-World Nature of Earthquake Network

The small-world nature is an important aspect of complex networks. It shows how a complex network is different from both regular and classical random graphs [8]. A small-world network resides in-between regularity and randomness, analogous to the edge of chaos in nonlinear dynamics.

To study the small-world nature of an earthquake network, a full network has to be reduced to a simple undirected graph: that is, loops are removed and each multiple edge is replaced by a single edge (see Fig. 1b). This is because in the small-world picture one is concerned only with simple linking pattern of vertices.

A small-world network is characterized by a large value of the clustering coefficient in Eq. (2) and a small value of the average path length [8]. The clustering coefficient quantifies the tendency of two neighboring vertices of a given vertex to be connected to each other. A small-world network has a large value of the clustering coefficient, whereas the value for the classical random graph is very small [2-4,8]: $C_{\text{random}} = \langle k \rangle / N \ll 1$, where $N$ and $\langle k \rangle$ are the total number of vertices and the average value of connectivity, respectively.

In Table 1, the results are presented for the clustering coefficient and the average path length [23,24]. One finds that the values of the clustering coefficient are in fact much larger than those of the classical random graphs and the average path length is short. Thus, the earthquake networks reduced to simple graphs exhibit important features of small-world network.

**VI. Hierarchical Structure**

As seen above, seismicity generates a scale-free and small-world network. To investigate complexity of earthquake network further, one may examine if it is hierarchically organized [25]. The hierarchical structure can be revealed by analyzing the clustering coefficient as a function of connectivity. The connectivity-dependent clustering coefficient, $c(k)$, is defined in Eq. (3). This quantifies the adjacency of two vertices connected to a vertex with connectivity, $k$, and gives information on hierarchical organization of a network.

In Fig. 3, the plots of $c(k)$ are presented [25]. As can be clearly seen, the clustering coefficient of the undirected simple earthquake network asymptotically follows the scaling law

$$c(k) \sim k^{-\beta} \qquad (7)$$

with a positive exponent $\beta$. This highlights hierarchical organization of the earthquake network.

Existence of the hierarchical structure is of physical importance. The earthquake network has growth with preferential attachment [2-4,7]. However, the standard preferential-attachment-model is known to fail to generate hierarchical organization [9]. To mediate between growth with preferential attachment and the presence of hierarchical organization, the concept of vertex deactivation has been introduced in the literature [26]. According to this concept, in the process of network growth, some vertices deactivate and cannot acquire new edges any more. This has a natural physical implication for an earthquake network: active faults may be deactivated through the process of stress release. In addition, also the fitness model [11] is known to generate hierarchical organization. This model generalizes the preferential attachment rule in such a way that not only connectivity but also "charm" of vertices (i.e., attracting a lot of edges) are taken into account. Seismologically, fitness is considered to describe intrinsic properties of faults such as geometric configuration and stiffness. Both of these two mechanisms can explain a possible origin of the complex hierarchical structure, by which relatively new vertices have chances to become hubs of the network. In the case of an earthquake network, it seems plausible to suppose that the hierarchical structure may be due to both deactivation and fitness.

A point of particular interest is that the hierarchical structure disappears if weak earthquakes are removed. For example, setting a lower threshold for earthquake magnitude, say $M_{th} = 3$, makes it difficult to observe the power-law decay of the clustering coefficient. This, in turn, implies that the hierarchical structure of an earthquake network is largely supported by weak shocks.

**VII. Mixing Property**

The scale-free nature, small-worldness, growth with preferential attachment and hierarchical organization all indicate that earthquake networks are very similar to other known networks, for example, the Internet. However, there is at least one point which shows an essential difference between the two. It is concerned with the mixing property, which is relevant to the concept of the nearest-neighbor average connectivity, $\bar{k}_{nn}(k)$ [in Eq. (4)], of a full network with loops and multiple edges.

The plots of this quantity are presented in Fig. 4. There, the feature of assortative mixing [25] is observed, since $\bar{k}_{nn}(k)$ increases with respect to connectivity *k*. Therefore, vertices with large values of connectivity tend to be linked to each other. That is, vertices containing stronger shocks tend to be connected among themselves with higher probabilities.

To quantify this property, the correlation coefficient in Eq. (5) is evaluated [25]. The result is summarized in Table 2. The value of the correlation coefficient is in fact positive, confirming that the earthquake network has assortative mixing. On the other hand, the Internet is of disassortative mixing [3,10-12]. That is, the mixing properties of the earthquake network and the Internet are opposite to each other. It is noticed however that the loops and multiple edges play essential roles for the assortative mixing: an undirected simple graph obtained by reducing a full earthquake network turns out to exhibit disassortative mixing. These are purely the phenomenological results, and their physical origins are still to be clarified.

## VIII. Period Distribution

So far, directedness of an earthquake network has been ignored. The full directed network picture is radically different from the small-world picture for a simple undirected graph. It enables one to consider interesting dynamical features of an earthquake network. As an example, here the concept of period [27] is discussed. This is relevant to the question "after how many earthquakes does an event return to the initial cell, statistically?" It is therefore of obvious interest for earthquake prediction.

Period in a directed network is defined as follow. Given a vertex of a network, there are various closed routes starting from and ending at this vertex. The period, $L_p$, of a chosen closed route is simply the number of edges forming the route (see Fig. 5).

The period distribution, $P(L_p)$, is defined as the number of closed routes. The result is presented in Fig. 6 [27]. As can be seen there, $P(L_p)$ obeys a power law

$$P(L_p) \sim (L_p)^{-\alpha}, \tag{8}$$

where $\alpha$ is a positive exponent. This implies that there exist a number of closed routes with significantly long periods in the network. This fact makes it highly nontrivial to statistically estimate the value of period.

## IX. Future Directions

In the above, the long-time statistical properties of an earthquake network have mainly been considered. On the other hand, given the cell size, an earthquake network represents all the dynamical information contained in a seismic time series, and therefore the study of its time evolution may give a new insight into seismicity. This, in turn, implies that it may offer a novel way of monitoring seismicity.

For example, it is of interest to investigate how the clustering coefficient changes in time as earthquake network dynamically evolves. According to the work in Ref. [28], the clustering coefficient remains stationary before a main shock, suddenly jumps up at the main shock, and then slowly decays to become stationary again following the power-law relaxation. In this way, the clustering coefficient successfully characterizes aftershocks in association with main shocks.

A question of extreme importance is if precursors of a main shock can be detected through monitoring dynamical evolution of earthquake network. Clearly, further developments are needed in science of complex networks to address to this question.

**Addendum**

In their article in this volume, Sornette and Werner make a criticism against the present complex-network description of seismicity simply based on a trivial point that no earthquake catalogs are complete. They ignore the fact that actually the effects of thresholding for magnitude on the properties of networks have carefully been examined and it has been ascertained that thresholding does not lead to significant changes in the structure of networks. It is a common sense in network science that the structure of a complex network has the high degree of robustness under "random attacks", i.e., random removal of vertices. The careful reader will never take Sornette and Werner's criticism serious.

# Figure and Table Captions

**Fig. 1a**

A schematic description of earthquake network. The dashed lines correspond to the initial and final events. The vertices, *A* and *B*, contain main shocks and play roles of hubs of the network.

**Fig.1b**

The undirected simple graph reduced from the network in Fig. 1a.

**Fig. 2**

The log-log plots of the connectivity distributions of the earthquake network constructed from the seismic data taken in California [the Southern California Earthquake Data Center (http://www.data.scec.org/)]. The time interval analyzed is between 00:25:8.58 on January 1, 1984 and 22:21:52.09 on December 31, 2003. The region covered is 29°06.00'N–38°59.76'N latitude and 113°06.00'W–122°55.59'W longitude with the maximal depth 175.99km. The total number of events is 367613. The data contains no threshold for magnitude (but "quarry blasts" are excluded from the analysis). Two different values of cell size are examined: (a) $10 \text{ km} \times 10 \text{ km} \times 10 \text{ km}$ and (b) $5\,\text{km} \times 5\,\text{km} \times 5\,\text{km}$. All quantities are dimensionless.

**Fig. 3**

The log-log plots of the connectivity-dependent clustering coefficient for two different values of cell size: (a) $10\,\text{km} \times 10\,\text{km} \times 10\,\text{km}$ and (b) $5\,\text{km} \times 5\,\text{km} \times 5\,\text{km}$. The analyzed period is between 00:25:8.58 on January 1, 1984 and 22:50:49.29 on December 31, 2004, which is taken from the same catalog as in Fig. 2. The region covered is 28°36.00'N–38°59.76'N latitude and 112°42.00'W–123°37.41'W longitude with the maximal depth 175.99km. The total number of the events is 379728. All quantities are dimensionless.

**Fig. 4**

The log-log plots of the nearest-neighbor average connectivity for two different values of cell size: (a) $10\,\text{km} \times 10\,\text{km} \times 10\,\text{km}$ and (b) $5\,\text{km} \times 5\,\text{km} \times 5\,\text{km}$. The data employed is the same as that in Fig. 3. The solid lines show the trends depicted by the exponentially increasing functions. All quantities are dimensionless.

**Fig. 5**

A full directed network: $\cdots \to v_1 \to v_2 \to v_3 \to v_2 \to v_2 \to v_4 \to v_3 \to v_2 \to \cdots$. The period associated with $v_3$ is 4, whereas $v_2$ has 1, 2 and 3.

**Fig. 6**

The log-log plots of the period distribution for two different values of cell size: (a) $10\,\text{km} \times 10\,\text{km} \times 10\,\text{km}$ and (b) $5\,\text{km} \times 5\,\text{km} \times 5\,\text{km}$. The data employed is the same as that in Fig. 2. All quantities are dimensionless.

**Table 1**

The small-world properties of the undirected simple earthquake network. The values of the number of vertices, *N*, the clustering coefficient, *C*, (compared with those of the classical random graphs, $C_{random}$) and the average path length, $L$ are presented. The data employed is the same as that in Fig. 2.

**Table 2**

The values of the dimensionless correlation coefficient. The data employed is the same as that in Fig. 3. Positivity of the values implies that mixing is assortative.

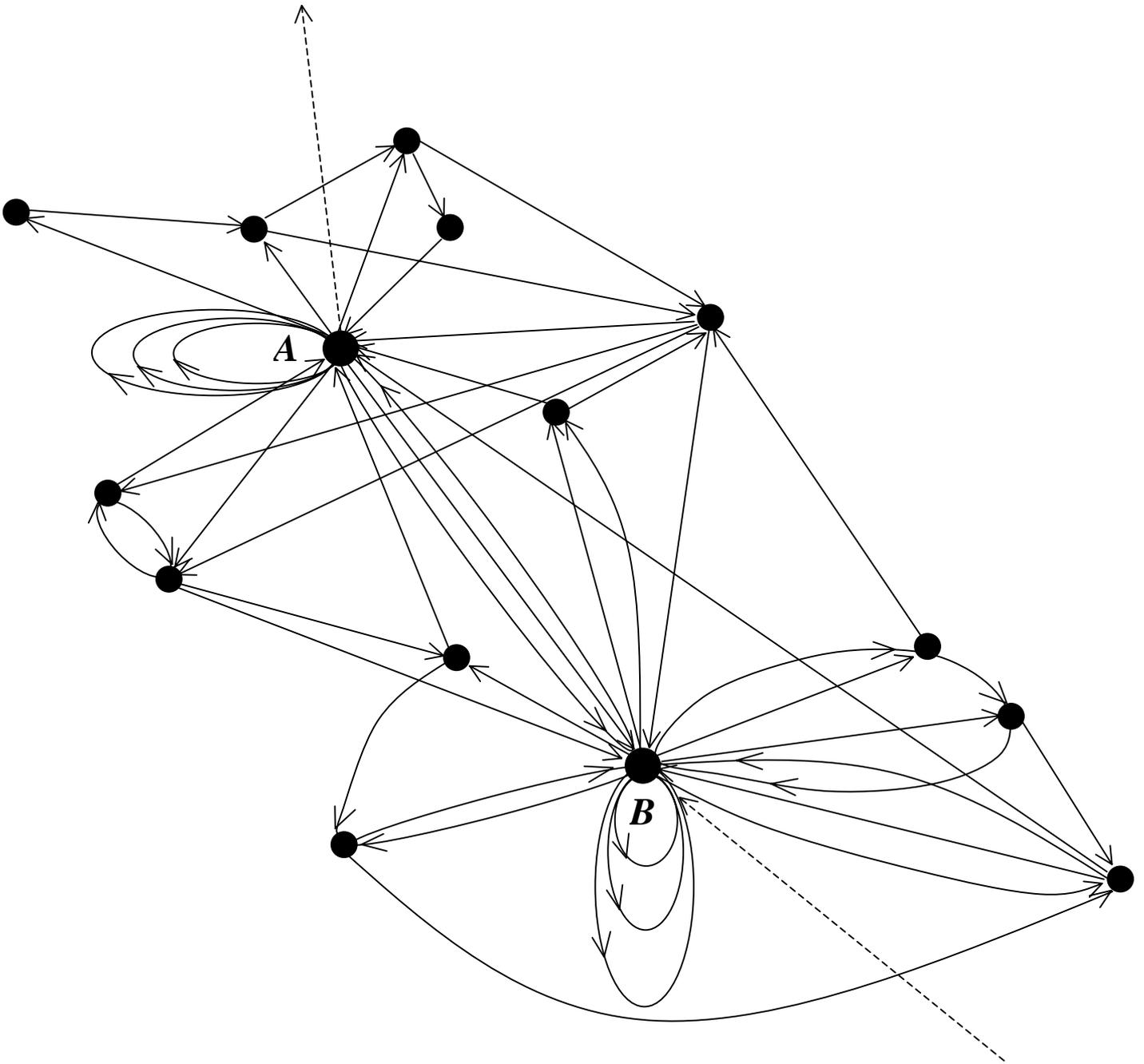

Fig.1a

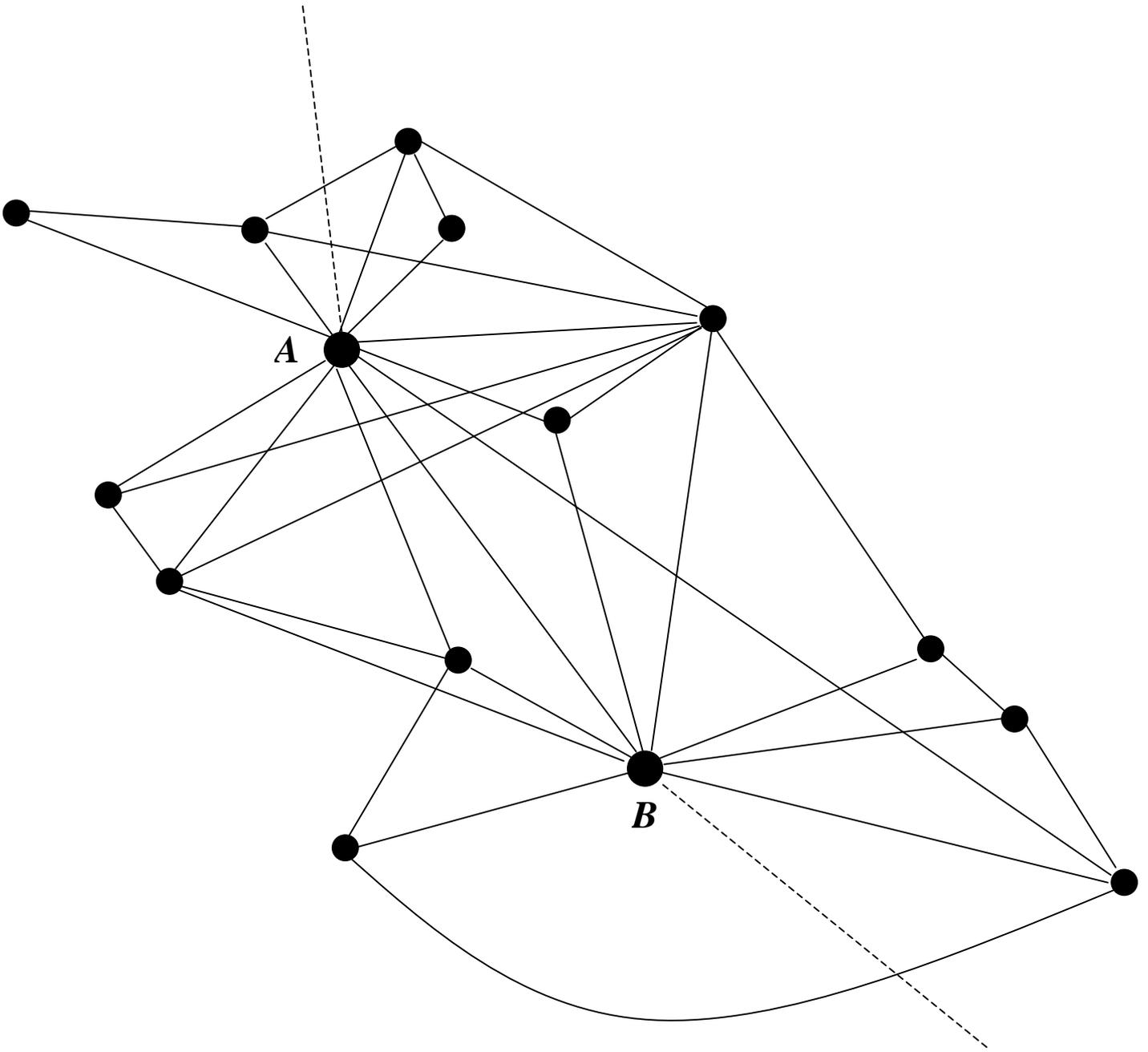

Fig.1b

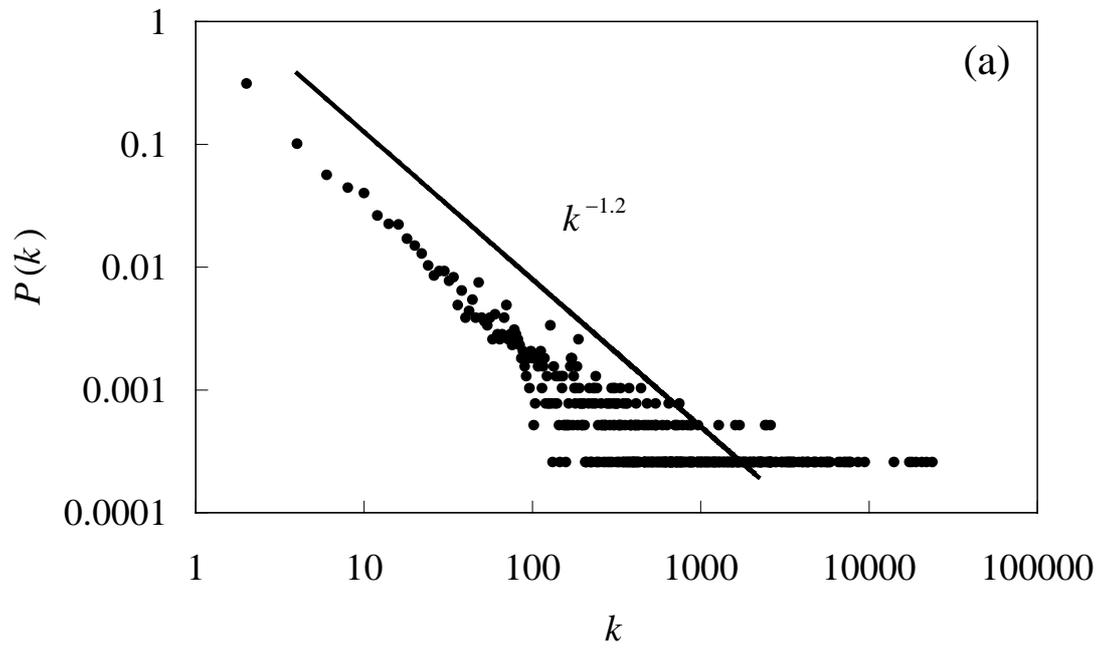

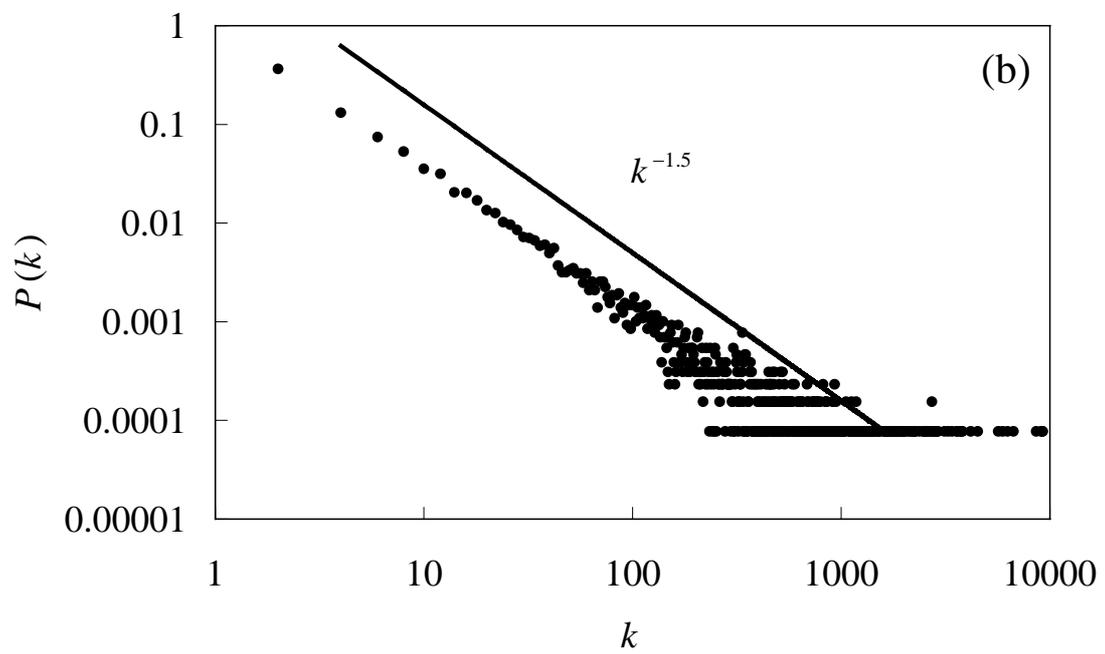

Fig.2

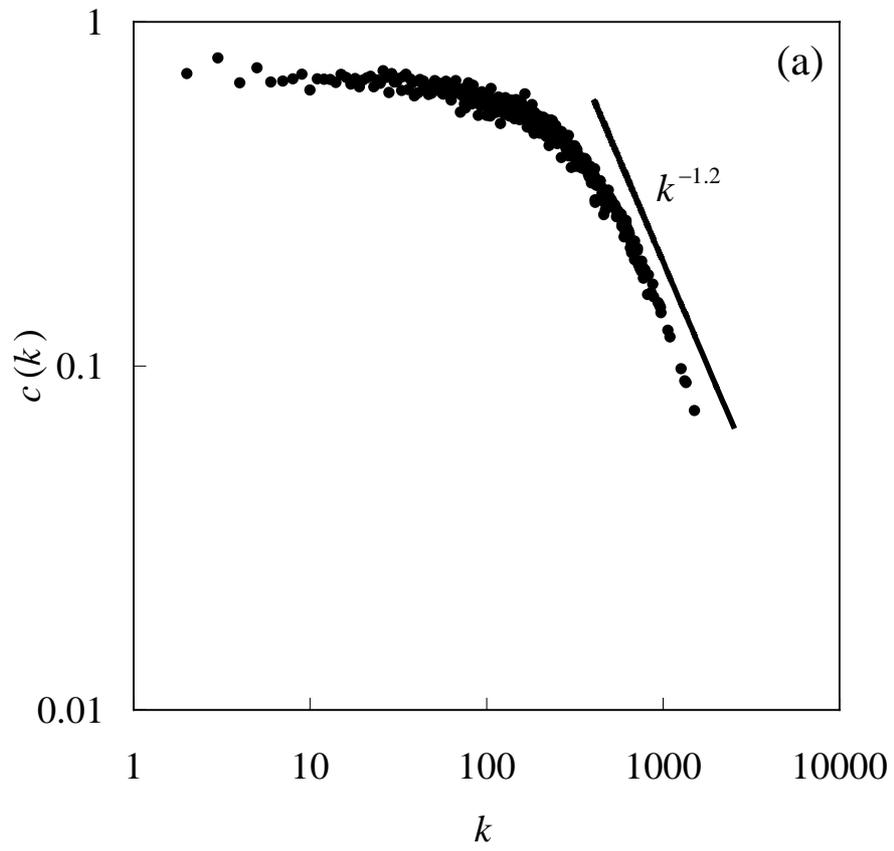

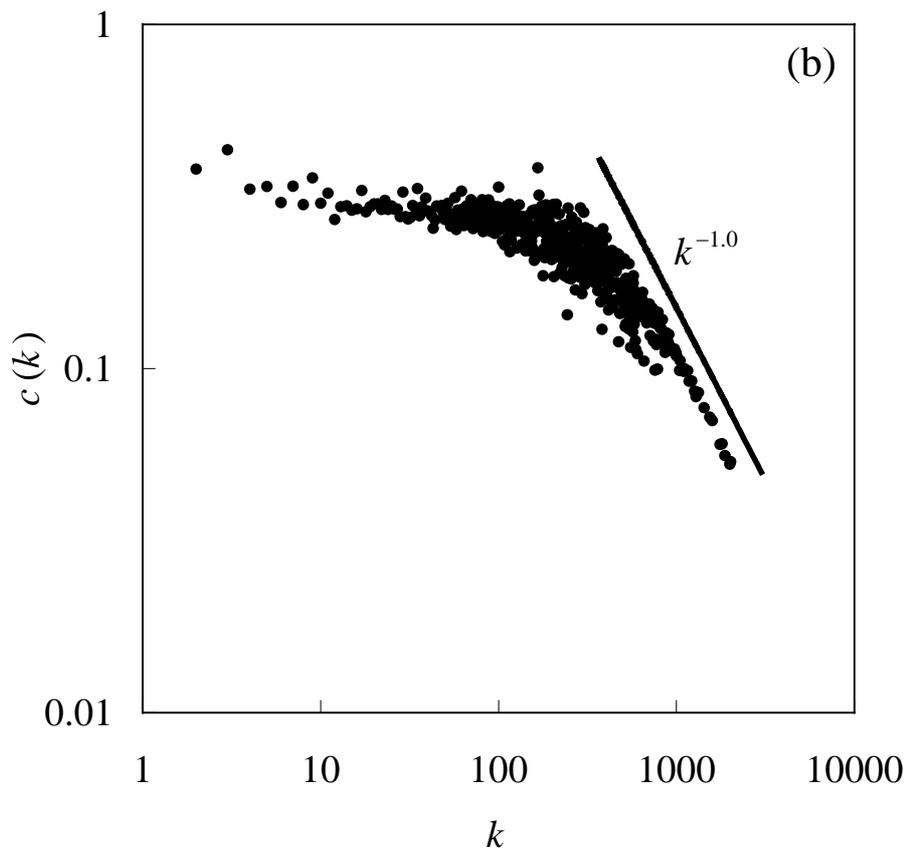

Fig.3

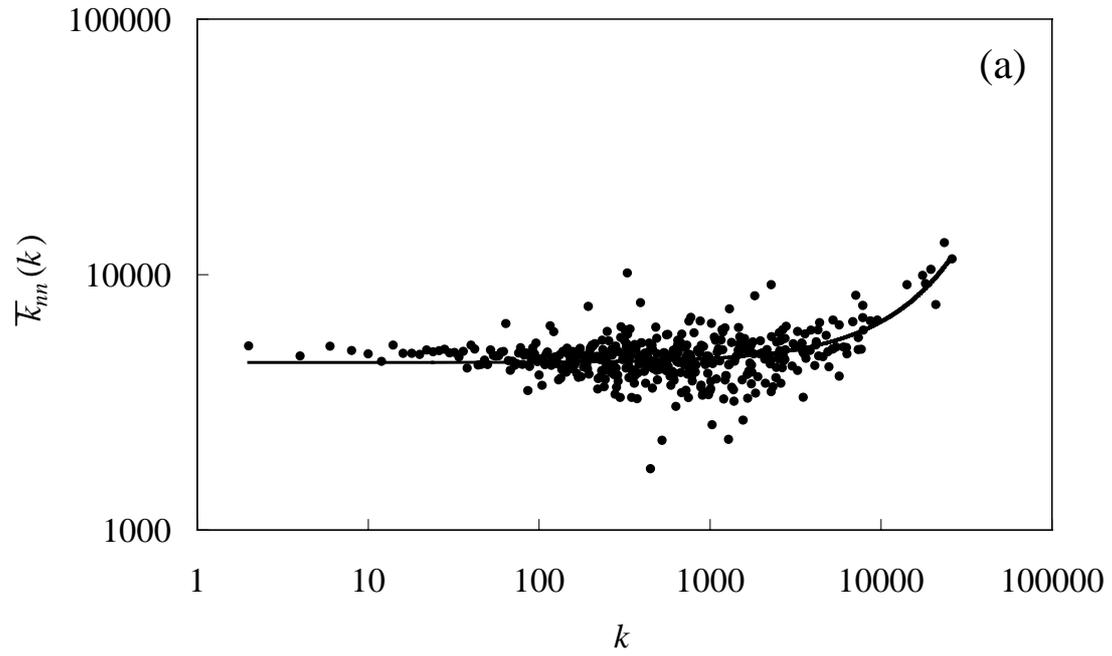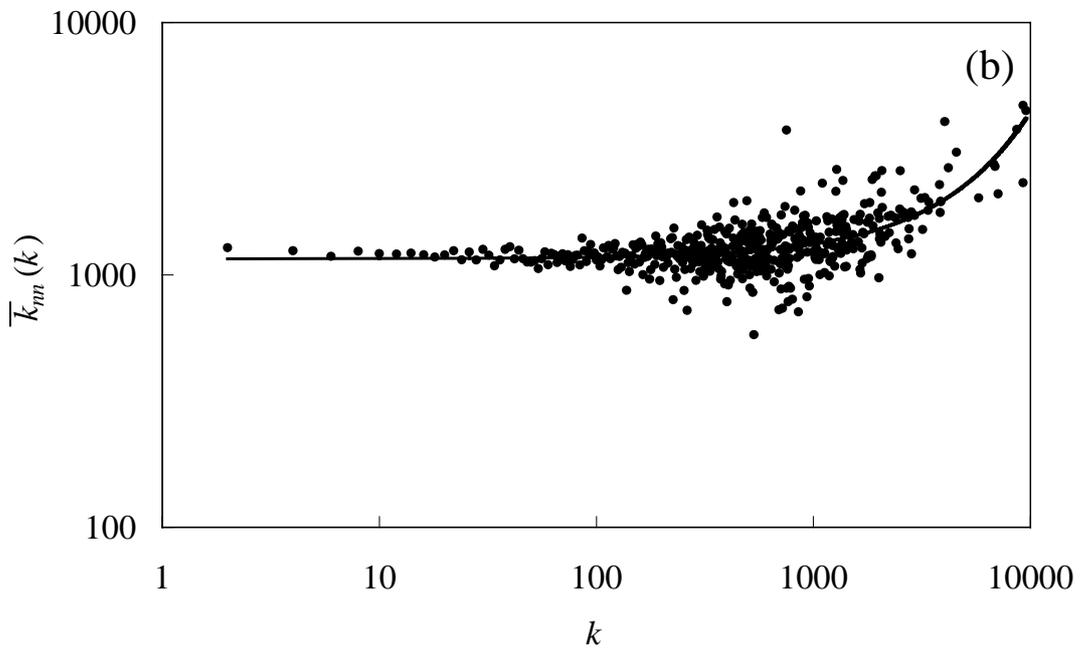

Fig.4

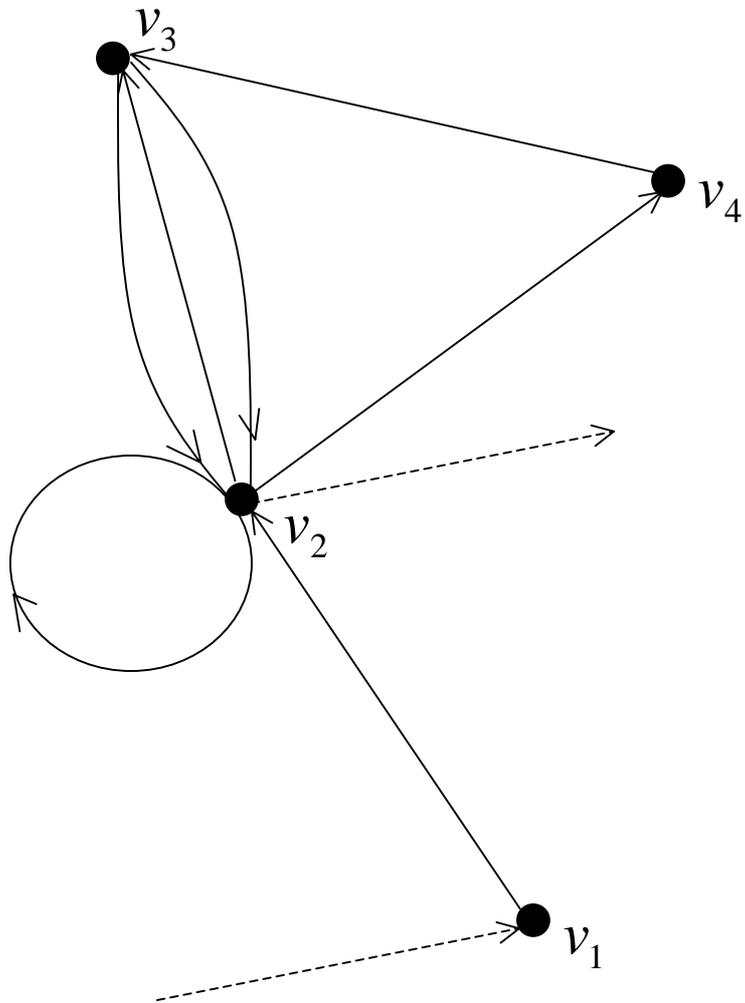

Fig.5

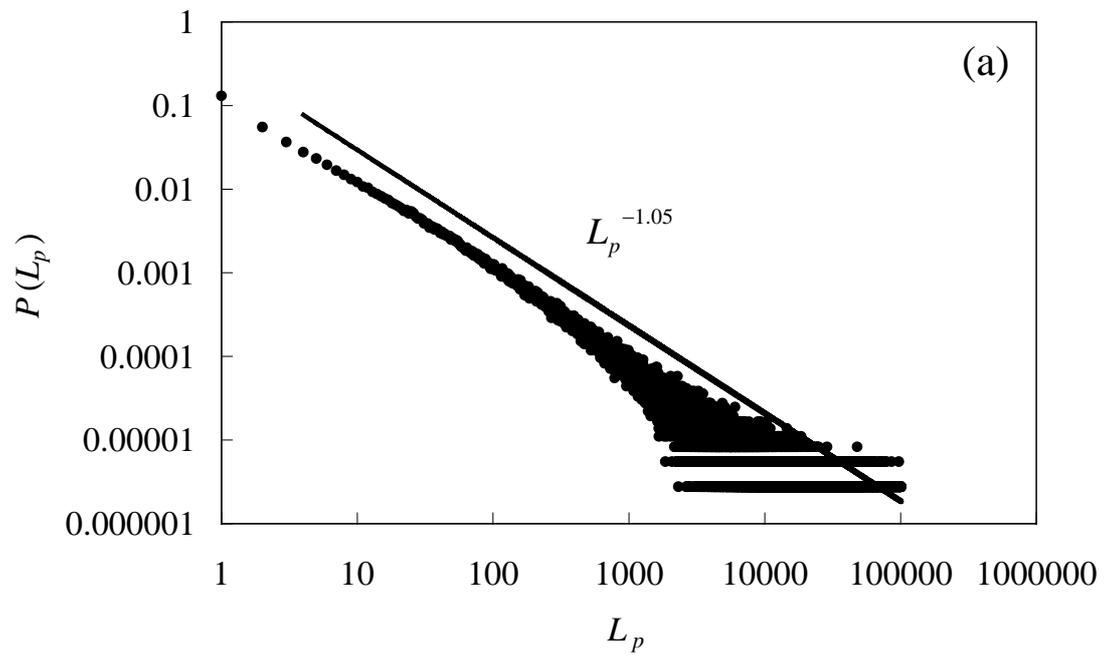

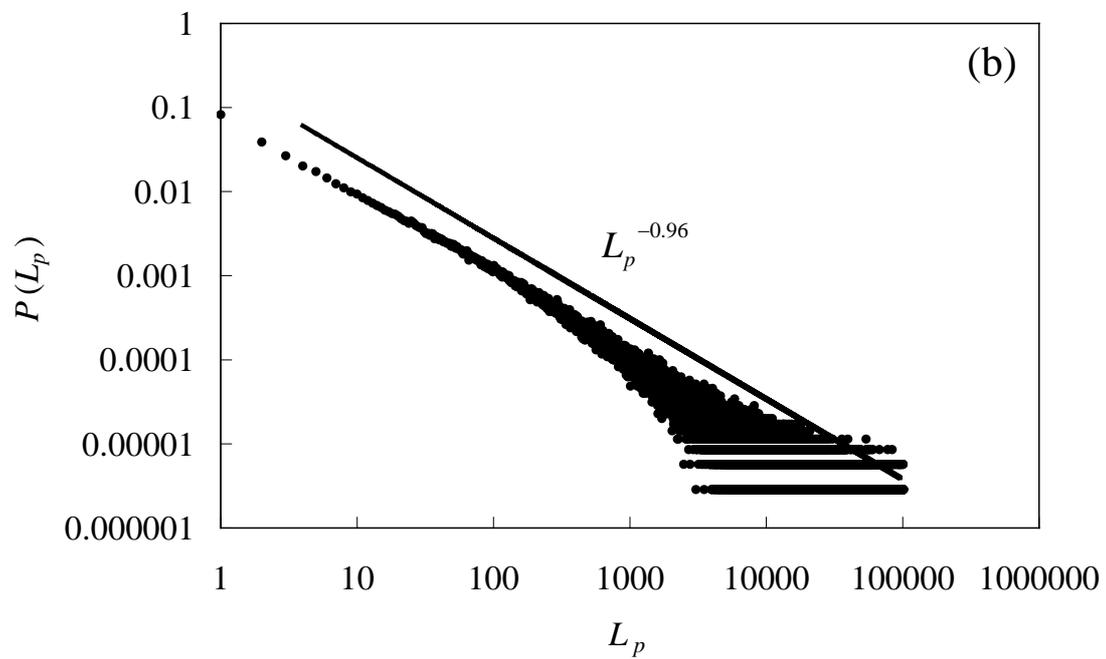

Fig.6

| cell size | $10\,\text{km} \times 10\,\text{km} \times 10\,\text{km}$ | $5\,\text{km} \times 5\,\text{km} \times 5\,\text{km}$ |
|---|---|---|
| number of vertices | $N = 3869$ | $N = 12913$ |
| clustering coefficient | $C = 0.630 \quad (C_{\text{random}} = 0.014)$ | $C = 0.317 \quad (C_{\text{random}} = 0.003)$ |
| average path length | $L = 2.526$ | $L = 2.905$ |

Table 1

| 10 km × 10 km × 10 km | 5 km × 5 km × 5 km |
|---|---|
| $r = 0.285$ | $r = 0.268$ |

Table 2